# Nonlinear Generation, Compression and Spatio-Temporal Analysis of GV/cm-Class Femtosecond Mid-Infrared Transients

C. Schoenfeld, L. Feuerer, A.-C. Heinrich, A. Leitenstorfer, and D. Bossini

*Department of Physics and Center for Applied Photonics, University of Konstanz, D-78457 Konstanz, Germany*

*\*Corresponding author: Davide.Bossini@uni-konstanz.de*

**A thin-disk regenerative amplifier with 3 kHz repetition rate pumps a second-order nonlinear mixing scheme providing femtosecond transients of maximum electric field strength beyond 330 MV/cm at a center frequency of 45 THz. This value surpasses field conditions present at sub-molecular dimensions of matter. The inherent competition between efficiency and bandwidth in parametric downconversion is overcome with a third-order nonlinear step of self-phase modulation. The high repetition rate, versatility, beam quality and passive phase stability of our system support advanced spectroscopic approaches like electro-optic sampling. With this technique, we study solitonic self-compression and octave-spanning supercontinua directly on a subcycle scale.**

___________________________________________________

Owing to a broad range of potential applications in both science and technology, generation of ultrashort and intense light pulses in the mid-infrared (MIR) spectral range is an attractive task. At the same time, it is demanding due to the absence of adequate gain media. For example, recent advances in this direction open up unprecedented possibilities in the realm of extreme and off-resonant optical biasing of condensed-matter systems [1–3]. Extension of the cutoff energy of gas-phase high-harmonics generation into the hard X-ray region is enabled by the quadratic scaling of the ponderomotive potential with the driving wavelength [4–6]. Many options exist for resonant nonlinear coupling as the eigenfrequencies of most collective excitations of charges, spins and the lattice of complex matter reside at MIR frequencies [6–10]. The hallmark peak intensities of laser systems required for extending these tasks are given by the Coulomb fields that arise at both the inter- and intra-atomic levels. They are on the scale of GV/cm, as determined by the elementary charges of electrons and protons occurring over Å-scale distances.

Recent efforts towards intensive access to the MIR region rely on near-infrared (NIR) pump sources combined with nonlinear optical steps for conversion to lower frequencies [10–18]. Second- and third-order processes are widely established but both exhibit shortcomings: Difference frequency generation (DFG) between two NIR pulses is limited by phase-matching constraints requesting a trade-off between acceptance bandwidth and efficiency due to the finite mismatch of group velocities [19]. On the other hand, self-phase modulation (SPM) in the NIR offers great bandwidth but wavelengths longer than 6 μm are out of reach [20].

Here, we solve the dilemma between efficiency and acceptance bandwidth by employing a combination of two separate nonlinear processes, namely second-order parametric down-conversion followed by spectral broadening based on SPM. This approach results in femtosecond MIR pulses close to GV/cm electric field strengths comparable with atomic and intra-molecular conditions. At the same time, central frequencies around 45 THz still allow for straightforward and sensitive subcycle analysis.

A scheme of the experiment is depicted in Fig. 1(a). A regenerative Yb:YAG thin-disk amplifier providing up to 17 mJ of close to transform-limited pulses of 615 fs duration at a center wavelength of 1030 nm (frequency of 291 THz) and repetition rate of 3 kHz with a beam quality of $M^2$ = 1.5 is seeded by a versatile Er/Yb:fiber frontend [21]. Half of this output serves as a pump pulse for subsequent DFG. A second branch is converted to lower frequencies by driving a mJ-class nonlinear optical parametric chirped-pulse amplifier (OPCPA) consisting of white-light generation in YAG and three consecutive amplification stages. Optimum temporal overlap between pump and signal is ensured by a pair of Brewster-angled ZnS wedges for dispersive control of the signal pulse duration after the first stage. Subsequently, both arms are individually expanded to a $1/e^2$ beam diameter of 6 mm and spatially overlapped for collinear type-II DFG in silver gallium sulfide ($AgGaS_2$, AGS) crystals with clear aperture of 8 mm. Passively phase-locked MIR pulses emerge from this approach. $AgGaS_2$ is selected due to its wide band gap of 2.8 eV [22] exceeding the photon energy of 1.2 eV of the pump at 291 THz by more than a factor of two. In this way, losses due to two-photon absorption are avoided. Fig. 1(b) displays the spectra of all three pulses involved in the nonlinear frequency conversion for two different configurations of OPCPA and $AgGaS_2$. For a more broadband configuration, we employ type-I β-barium borate (BBO) crystals in all OPCPA stages along with a 0.15 mm-thin $AgGaS_2$ emitter for DFG. The field traces resulting from electro-optic sampling (EOS) are depicted in Fig. 2(a). Fast Fourier transform reveals amplitude spectra spanning from below 30 to above 55 THz (Fig. 2(b)). The bright red data points in Fig. 1(c) illustrate the MIR pulse energies as a function of pump energy for the DFG. A maximum output energy of 380 nJ is achieved pumping with 2 mJ, corresponding to an energy conversion efficiency of 0.02 %. On the other hand, generation of narrowband and highly intense MIR pulses involves type-II BBO crystals in the first two OPA stages and type-I phase matching in the third one, combined with 2 mm-thick $AgGaS_2$. Here, the highest pulse energy (dark red graph in Fig. 1(c)) amounts to 49 µJ, corresponding to an energy conversion efficiency of 2.7 %. Note that in this case almost 20 % of pump photons underwent down-conversion during DFG. Degradation of the $AgGaS_2$ surface is observed when the pump energy exceeds 2 mJ, corresponding to a damage threshold of 14 mJ/cm$^2$.

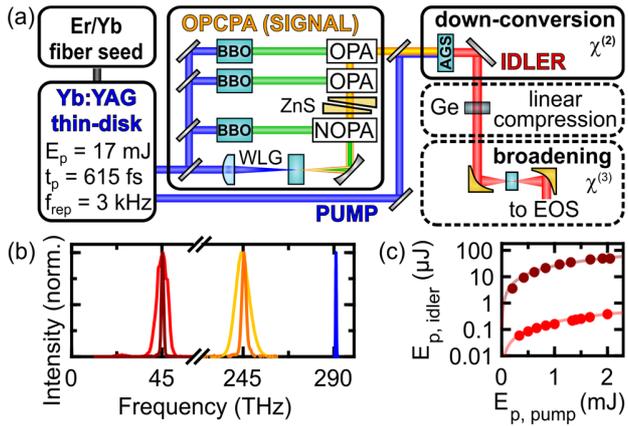

**Figure 1.** (a) Scheme of the experimental setup consisting of an ultrabroadband Er/Yb:fiber frontend, a regenerative sub-ps Yb:YAG thin-disk amplifier, an optical parametric chirped pulses amplifier (OPCPA) seeded by white-light generation (WLG), followed by multi-THz difference frequency generation (DFG) and compression units. (b) Spectra of the regenerative Yb:YAG thin-disk amplifier (pump; blue), narrow- and broadband examples of OPCPA emission (signal; orange and yellow, respectively) and DFG output (idler; dark red and red graphs) generated in $AgGaS_2$ (AGS). (c) MIR pulse energy versus near-infrared pump pulse energy for $AgGaS_2$ crystals with 0.15 mm (red) and 2 mm thickness (dark red). Solid lines represent quantum efficiencies of 0.12 % and 20 %, respectively.

Since the EOS traces in Fig. 2 have to be taken with strongly attenuated field amplitudes far out of focus, absolute determination of the confocal electric fields requires exact knowledge of energy densities both in space and time. Under our conditions, total pulse energies readily follow from the average power measured with a thermopile detector and the repetition rate. Relative intensity envelopes may be obtained by squaring the temporal electric field traces from EOS. Depending on the OPCPA configuration and thickness of the $AgGaS_2$ crystal employed, either few-cycle and broadband (Figs. 2(a,b)) or narrowband multicycle pulses (Figs. 2(d,e)) emerge. Note that the positive chirp of the signal set by the ZnS wedges is designed to result in negatively chirped MIR pulses after DFG in the $AgGaS_2$ crystal. Therefore, the wave packets may be controlled by simple propagation through a medium with normal dispersion such as germanium. Pulse durations corresponding to the broadband amplitude spectrum in Fig. 2(b) are depicted versus thickness of antireflection-coated Ge windows used for compression in Fig. 2(c). The flat phase spectrum in Fig. 2(b) indicates a transform-limited transient with a full-width-at-half-maximum (FWHM) duration of the intensity envelope of 67 fs obtained for a Ge thickness of 13 mm (Fig. 2(a)). Note that the narrowband pulses generated in 2 mm-thick $AgGaS_2$ (Figs. 2(d,e)) directly emerge close to the transform limit and are hardly affected by the Ge insertion. The second information necessary to calculate the electric field amplitude is the confocal intensity distribution of the MIR beam. Ideally, focusing should be as tight as possible since the peak electric field strength scales inversely with the beam waist. We determine this quantity with the knife-edge method after attenuating with a neutral-density filter to prevent ablation of the razor blade. The data set in Fig. 2(f) yields a beam diameter of 11 µm (FWHM of the intensity) focusing with an off-axis parabolic mirror of a focal length of 5 mm. This value is close to the diffraction limit and corresponds to $M^2 = 1.7$. We emphasize that, despite the massive amount of nonlinear optical interactions employed in our setup, this value remains remarkably close to the beam quality of the near-infrared pump. The inset of Fig. 2(f) shows a Gaussian-like and circular transverse mode profile in the far field, as taken with an infrared beam profiling camera (Ophir Sphiricon - Pyrocam IIIHR). Considering the exact temporal waveforms and the spatial energy densities, we calculate a peak amplitude of the electric field of 330 MV/cm for the narrowband multicycle configuration and of

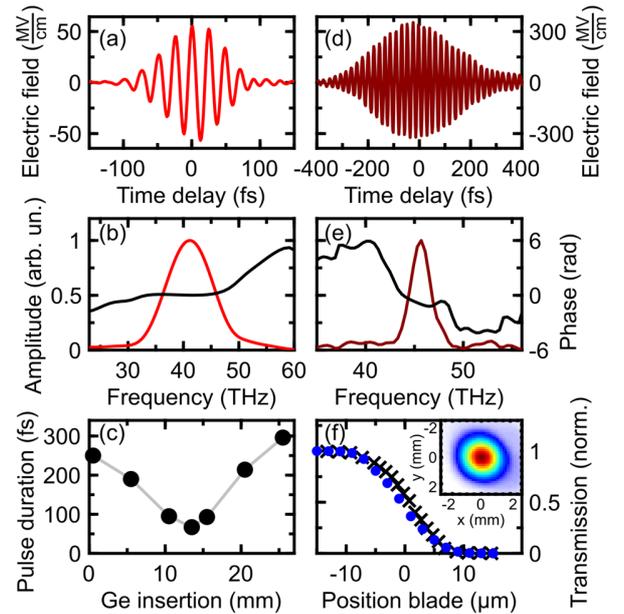

**Figure 2.** (a) Electric-field trace of a MIR pulse generated via DFG in 0.15 mm-thick $AgGaS_2$, temporally compressed via propagation through a 13 mm-thick Ge window and focused to a spot size of 11 µm. (b) Corresponding amplitude (red) and phase (black) spectra obtained via Fourier transform. (c) MIR pulse duration as a function of the thickness of Ge insertion. (d) Electric-field transient of a multicycle MIR pulse generated via DFG in a 2 mm-thick $AgGaS_2$ and focused to a spot size of 11 µm. (e) Amplitude and phase spectra corresponding to the time-domain data in (d). (f) Transmitted intensity versus position of a knife edge translated along the vertical (blue dots) and horizontal axis (black crosses) for the transient depicted in (d). The resulting spot size is 11 µm (FWHM of the intensity). Inset shows the transverse intensity profile in the far field.

56 MV/cm for the broadband sub-three-cycle operation, respectively. Notably, a value of 330 MV/cm corresponds to the Coulomb field at a distance of 2 Å from an elementary point charge (1.6·10$^{-19}$ C), comparable e.g. to the atomic radius of elements such as Sr or La [23] or to the bond length of typical molecules such as $Cl_2$ [24]. Still, our results exploiting second-order conversion suffer from the incompatibility of efficiency and acceptance bandwidth in the DFG process. Assuming no pump depletion, the output energy increases exponentially with the length of the nonlinear medium while the bandwidth decreases with the inverse square root owing to the finite mismatch of group velocities [19]. Here, we overcome this issue by first generating narrowband and intense mid-IR pulses, see Figs. 2(d,e). We then increase the bandwidth via third-order spectral broadening and solitonic compression. This step requires materials with a broad window of transparency and negative group velocity dispersion (GVD) which are readily available in the MIR. Promising candidates include thallium bromo-iodide (KRS-5), CsI, ZnSe, KBr and NaCl with corresponding zero-dispersion frequencies of 45 THz, 52 THz, 57 THz, 78 THz and 109 THz, as calculated from Sellmeier equations [25–27]. The multicycle pulses are focused into the anomalously dispersive sample and recollimated by off-axis parabolic mirrors with focal lengths of 50 mm, yielding a focus spot size of 70 µm (FWHM diameter, see Fig. 1(a)). The MIR fluence is systematically varied to control SPM and to balance its interplay with the anomalous dispersion. All details of the nonlinear pulse propagation, including the solitonic dynamics that arises, are directly accessible on a subcycle level of the electric fields via EOS. Figure 3(a) shows the evolution of the field transients and intensity profiles for increasing MIR fluences in a 4 mm-thick KRS-5 crystal with AR coating. Corresponding amplitude spectra are depicted in Fig. 3(b). For comparison, the black graphs denote the conditions in front of the KRS-5 crystal. At a peak fluence of 43 mJ/cm$^2$, clear evidence of nonlinear dynamics such as pulse splitting and the onset of spectral broadening is observed. Temporal distortions indicate that chromatic dispersion and soliton fission play a pivotal role in the broadening process. By further increasing the MIR fluence and thus the influence of SPM effects, a temporal shock front with rise time shorter than 100 fs arises because of self-steepening [28]. Beyond this regime, the FWHM of the intensity profile experiences a significant reduction to 100 fs, corresponding to a temporal compression by a factor of four. In comparison with the reference transient, we obtain an increase in peak field and intensity by factors of 1.3 and 1.7 through this nonlinear compression scheme, respectively. Note that this result is achieved despite 40 % of total energy losses owing to reflection and nonlinear absorption. Compared to the broadband pulses generated directly by thin $AgGaS_2$ crystals, the field enhancement achieved with nonlinear compression in KRS-5 amounts to 400 % at comparable bandwidths. Consequently, our approach of combining second- and third-order nonlinear processes to efficiently generate and compress MIR pulses overcomes the limitations associated with the phase mismatch of the DFG process, effectively providing GV/cm field amplitudes with sub-100 fs pulse duration. We note that a minimum pulse duration together with maximum field amplitudes is mandatory for e.g. optimal biasing of condensed matter into exotic states like Wannier-Stark localization because detrimental effects such as off-resonant generation and acceleration of hot charge carriers are minimized. We also observe no signs of long-term degradation in KRS-5, consistent with the literature [29]. In contrast, alkali halides such as NaCl, KBr and CsI support formation of color centers for fluences higher than 43 mJ/cm$^2$, severely restricting use of these materials [30]. To ensure long-term stable operation, we increase the focal length of the off-axis parabola to 100 mm, effectively reducing the confocal intensity by a factor of four. Figs. 3(c) and (d) show the field traces and spectral evolution in 20 mm of KBr. Also in this case, we observe significant spectral broadening but without temporal compression due to a finite misbalance between SPM effects and anomalous dispersion. Fig. 4(a) summarizes the spectral broadening in KRS-5, CsI, ZnSe, KBr and NaCl for increasing fluence. The largest bandwidth increase by a factor of 6 is observed with a 20 mm-thick KBr crystal. For SPM-dominated spectral broadening, the nonlinear refractive index $n_2$ may be determined from

$$\Delta f \approx \frac{f_0 \, P_{max} \, z}{c t_p A} n_2 \qquad (1)$$

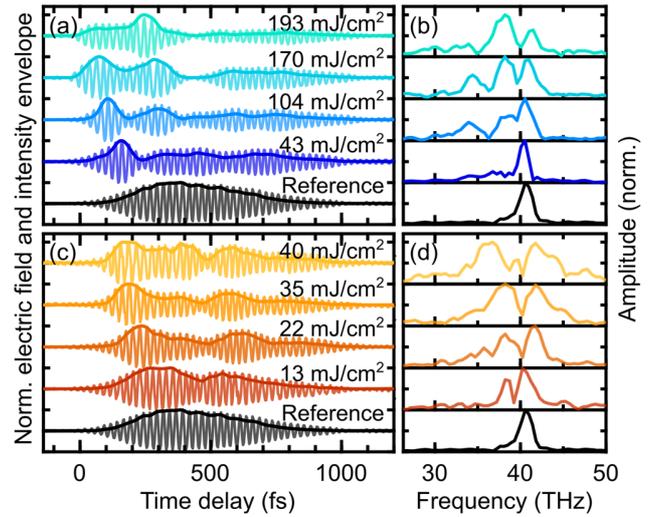

**Figure 3.** Field-resolved investigation of nonlinear spectral broadening in anomalously dispersive media: Electric-field waveforms (shaded lines) and intensity profile (solid lines), as obtained by electro-optic sampling for increasing incident fluences and after propagation through (a) 4 mm of thallium bromo-iodide (KRS-5) and (c) 20 mm of KBr. (b) and (d): Corresponding amplitude spectra, as obtained directly from the time-domain traces via Fourier transform, respectively.

[31] where $\Delta f$ describes the resulting spectral bandwidth, $f_0$ the central frequency of incident light, c the speed of light, $P_{max}$ the peak power, $A$ the cross-sectional area at the beam waist, $t_p$ the pulse duration and $z$ the interaction length which corresponds to twice the Rayleigh range for the case of free-space focusing. Exploiting Eq. (1), we obtain $n_{2,\,KBr}$ = 3·10$^{-15}$ cm$^2$/W and $n_{2,\,NaCl}$ = 0.9·10$^{-15}$ cm$^2$/W, good in agreement with existing literature [30]. A saturation of the spectral broadening is observed in Fig. 4(a) except for KBr and NaCl. Several effects may contribute to this effect: (i) nonlinear absorption related to generation of inverse bremsstrahlung [32] or the dynamic Franz-Keldysh effect [33], (ii) limited sensitivity of EOS to frequencies above 55 THz due to the finite bandwidth of our probe pulses and (iii) geometrical aspects of the third-order interaction like self-focusing and filamentation. To examine the third point, we study the spatial evolution of the transverse intensity profile in the far field after focusing through KBr and KRS-5 for increasing MIR fluences (Fig. 4(b)). Only minute

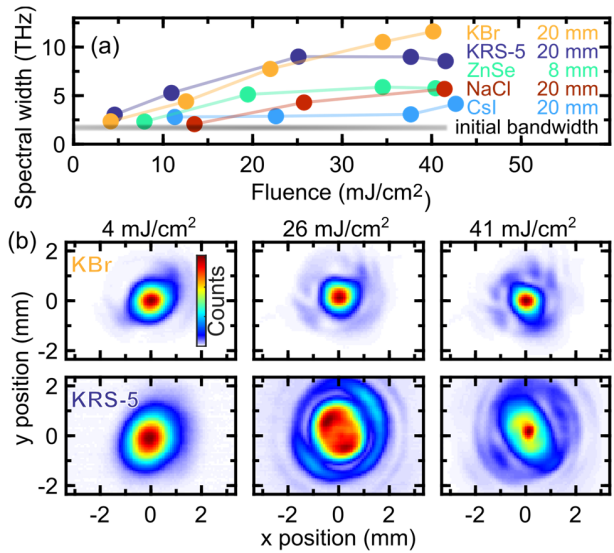

**Figure 4.** (a) Spectral broadening (1/e width) as a function of incident fluence for CsI, KBr, NaCl, ZnSe and KRS-5 crystals of various thicknesses. The focal length used is 100 mm, resulting in a confocal beam diameter of 155 μm and a Rayleigh length of 8 mm. (b) Transverse mode profile in far field after propagation through KBr and KRS-5 with thickness of 20 mm for increasing incident fluence, as taken with an infrared beam profiling camera.

changes are observed in KBr. In contrast, multi-filamentation appears at 26 mJ/cm$^2$ in KRS-5, i.e. at the same intensity where the spectral broadening saturates. This break-up of the transverse mode into many filaments arises due to inhomogeneities in the sample as well as imperfections of the beam profile. If the fluence is further increased, the wave front collapses and a hot-spot mode emerges at 41 mJ/cm$^2$.

In conclusion, we have presented a system for generation of intense MIR pulses down to the few-cycle regime which combines the high average power supplied by Yb:YAG thin-disk technology with a multi-stage setup for parametric amplification and DFG. Output pulses of a central frequency of 45 THz reach peak electric field amplitudes of 56 MV/cm in a broadband sub-three-cycle regime or 330 MV/cm in a multi-cycle configuration, respectively. These values are comparable to intra-atomic conditions. The trade-off between bandwidth and efficiency for second-order frequency conversion is avoided by third-order spectral broadening. The passive phase stability and versatility of our setup allows subcycle analysis of the electric field traces via EOS. Consequently, our light source enables a new regime of sensitive investigations of extremely nonlinear and nonperturbative light-matter interactions both under resonant driving of collective resonances and purely off-resonant biasing.

**Funding.** Deutsche Forschungsgemeinschaft (SFB 1432, 425217212), Davide Bossini acknowledges the support of the DFG program BO 5074/1-1.

**Acknowledgments** The authors thank A. Baserga, C. Beschle and S. Eggert for their technical support.

**Disclosures.** The authors declare no conflicts of interest.

**Data availability**. Data underlying the results presented in this paper are not publicly available at this time but may be obtained from the authors upon reasonable request.

1. O. Schubert, M. Hohenleutner, F. Langer, B. Urbanek, C. Lange, U. Huttner, D. Golde, T. Meier, M. Kira, S. W. Koch, and R. Huber, Nature Photon **8**, 119 (2014).
2. C. Schmidt, J. Bühler, A.-C. Heinrich, J. Allerbeck, R. Podzimski, D. Berghoff, T. Meier, W. G. Schmidt, C. Reichl, W. Wegscheider, D. Brida, and A. Leitenstorfer, Nat Commun **9**, 2890 (2018).
3. D. Berghoff, J. Bühler, M. Bonn, A. Leitenstorfer, T. Meier, and H. Kim, Nat Commun **12**, 5719 (2021).
4. B. Shan and Z. Chang, Phys. Rev. A **65**, 011804 (2001).
5. T. Popmintchev, M.-C. Chen, D. Popmintchev, P. Arpin, S. Brown, S. Ališauskas, G. Andriukaitis, T. Balčiunas, O. D. Mücke, A. Pugzlys, A. Baltuška, B. Shim, S. E. Schrauth, A. Gaeta, C. Hernández-García, L. Plaja, A. Becker, A. Jaron-Becker, M. M. Murnane, and H. C. Kapteyn, Science **336**, 1287 (2012).
6. G. Vampa, T. J. Hammond, N. Thiré, B. E. Schmidt, F. Légaré, C. R. McDonald, T. Brabec, and P. B. Corkum, Nature **522**, 462 (2015).
7. M. Först, C. Manzoni, S. Kaiser, Y. Tomioka, Y. Tokura, R. Merlin, and A. Cavalleri, Nature Phys **7**, 854 (2011).
8. T. F. Nova, A. Cartella, A. Cantaluppi, M. Först, D. Bossini, R. V. Mikhaylovskiy, A. V. Kimel, R. Merlin, and A. Cavalleri, Nature Phys **13**, 132 (2017).
9. A. Stupakiewicz, C. S. Davies, K. Szerenos, D. Afanasiev, K. S. Rabinovich, A. V. Boris, A. Caviglia, A. V. Kimel, and A. Kirilyuk, Nat. Phys. **17**, 489 (2021).
10. R. Huber, A. Brodschelm, F. Tauser, and A. Leitenstorfer, Applied Physics Letters **76**, 3191 (2000).
11. A. Sell, A. Leitenstorfer, and R. Huber, Opt. Lett., OL **33**, 2767 (2008).
12. D. Brida, C. Manzoni, G. Cirmi, M. Marangoni, S. Bonora, P. Villoresi, S. D. Silvestri, and G. Cerullo, J. Opt. **12**, 013001 (2009).
13. G. Andriukaitis, T. Balčiūnas, S. Ališauskas, A. Pugžlys, A. Baltuška, T. Popmintchev, M.-C. Chen, M. M. Murnane, and H. C. Kapteyn, Opt. Lett., OL **36**, 2755 (2011).
14. C. Gaida, M. Gebhardt, T. Heuermann, F. Stutzki, C. Jauregui, J. Antonio-Lopez, A. Schülzgen, R. Amezcua-Correa, A. Tünnermann, I. Pupeza, and J. Limpert, Light Sci Appl **7**, 94 (2018).
15. U. Elu, L. Maidment, L. Vamos, F. Tani, D. Novoa, M. H. Frosz, V. Badikov, D. Badikov, V. Petrov, P. St. J. Russell, and J. Biegert, Nat. Photonics **15**, 277 (2021).
16. M. Seidel, X. Xiao, S. A. Hussain, G. Arisholm, A. Hartung, K. T. Zawilski, P. G. Schunemann, F. Habel, M. Trubetskov, V. Pervak, O. Pronin, and F. Krausz, Science Advances **4**, eaaq1526 (2018).
17. O. Novák, P. R. Krogen, T. Kroh, T. Mocek, F. X. Kärtner, and K.-H. Hong, Opt. Lett., OL **43**, 1335 (2018).
18. W. Harmon, K. Robben, and C. M. Cheatum, Opt. Lett., OL **48**, 4797 (2023).
19. C. Manzoni and G. Cerullo, J. Opt. **18**, 103501 (2016).
20. K. L. Vodopyanov, (John Wiley & Sons, 2020).
21. J. Fischer, A.-C. Heinrich, S. Maier, J. Jungwirth, D. Brida, and A. Leitenstorfer, Opt. Lett. **41**, 246 (2016).
22. F. Rotermund, V. Petrov, and F. Noack, Optics Communications (2000).
23. J. C. Slater, Journal of Chemical Physics **41**, 3199 (1964).
24. D. R. Lide, (CRC Press, 2004).
25. W. S. Rodney and I. H. Malitson, J. Opt. Soc. Am., JOSA **46**, 956 (1956).
26. H. H. Li, Journal of Physical and Chemical Reference Data **5**, 329 (1976).
27. T. Amotchkina, M. Trubetskov, D. Hahner, and V. Pervak, Appl. Opt., AO **59**, A40 (2020).
28. G. P. Agrawal, Phys. Rev. A **44**, 7493 (1991).
29. A. Marcinkevičiūtė, G. Tamošauskas, and A. Dubietis, Optical Materials **78**, 339 (2018).
30. A. Marcinkevičiūtė, V. Jukna, R. Šuminas, N. Garejev, G. Tamošauskas, and A. Dubietis, Results in Physics **14**, 102396 (2019).
31. R. W. Boyd, in *Nonlinear Optics (Third Edition)*, R. W. Boyd, ed. (Academic Press, 2008), pp. 329–390.
32. A. Couairon and A. Mysyrowicz, Physics Reports **441**, 47 (2007).

33. A. H. Chin, J. M. Bakker, and J. Kono, Phys. Rev. Lett. **85**, 3293 (2000).